\documentclass[twocolumn,preprintnumbers,superscriptaddress,nofootinbib,aps,prd,floatfix]{revtex4}

\usepackage{enumerate}
\usepackage{subfigure}
\usepackage{amsmath}
\usepackage{amssymb}
\usepackage{graphicx}
\usepackage{placeins}
\usepackage{xspace,slashed}
\usepackage{hyperref}
\hypersetup{colorlinks=true, citecolor=blue, urlcolor=blue, linkcolor=blue}
\usepackage[normalem]{ulem}
\usepackage{booktabs,bbm}

\newcommand{\red}{\color{red}}

\begin{document}

\providecommand{\abs}[1]{\lvert#1\rvert}

\preprint{SI-HEP-2022-15}
\preprint{P3H-22-070}
\title{
 \boldmath MUonE, muon $g-2$ and electroweak precision constraints within 2HDMs}

\begin{abstract}
Two Higgs doublet models are attractive scenarios for physics beyond the
Standard Model. In particular, lepton-specific manifestations remain
contenders to explain the observed discrepancy between the anomalous
magnetic moment of the muon $a_\mu$ predicted within the Standard Model 
and recent observations at Fermilab and BNL. 
Dominant uncertainties that affect $a_\mu$ have motivated the MUonE
experiment to access the hadronic vacuum polarisation contribution that
impacts $a_\mu$ via elastic muon-electron scattering. 
In this work, we contrast the high precision that is achievable within the
MUonE context 
with 
constraints from flavour physics, 
precision electroweak constraints 
and LHC searches as well as their extrapolations 
for a range of two Higgs doublet models with a
softly broken $\mathbb{Z}_2$ symmetry. 
We find that the sensitivity of MUonE does not extend beyond the 
parameter regions that are already excluded by other constraints. 
MUonE will therefore provide a detailed measurement of the hadronic vacuum
polarisation contribution which then transparently informs $a_\mu$
interpretations in 2HDMs without modifications of correlations from beyond
the Standard Model interactions. 
In passing we extend earlier results of LHC and flavour projections to lepton-specific 2HDM (Types X and Y) scenarios, and comment on the possibility of modifying the value of the W-boson mass; 
we briefly discuss the implications for a strong first-order electroweak phase transition for these models.
\end{abstract}

\author{Oliver Atkinson} \email{o.atkinson.1@research.gla.ac.uk}
\affiliation{School of Physics and Astronomy, University of Glasgow, Glasgow G12 8QQ, United Kingdom\\[0.1cm]}
\author{Matthew Black} \email{Matthew.Black@uni-siegen.de}
\affiliation{Physik Department, Universit\"{a}t Siegen, Walter-Flex-Str. 3, 57068 Siegen, Germany\\[0.1cm]}
\author{Christoph Englert} \email{christoph.englert@glasgow.ac.uk}
\affiliation{School of Physics and Astronomy, University of Glasgow, Glasgow G12 8QQ, United Kingdom\\[0.1cm]}
\author{Alexander Lenz} \email{Alexander.Lenz@uni-siegen.de}
\affiliation{Physik Department, Universit\"{a}t Siegen, Walter-Flex-Str. 3, 57068 Siegen, Germany\\[0.1cm]}
\author{Aleksey Rusov}\email{rusov@physik.uni-siegen.de} 
\affiliation{Physik Department, Universit\"{a}t Siegen, Walter-Flex-Str. 3, 57068 Siegen, Germany\\[0.1cm]}

\maketitle

\section{Introduction}
\label{sec:intro}
Searches for new physics at the current high energy collider frontier of the Large Hadron Collider (LHC) have not revealed
any significant sign of new interactions beyond the Standard Model (BSM). This stands in contrast to some tensions
in precision analyses in flavour physics, see e.g. Refs.~\cite{Albrecht:2021tul,Gershon:2021pnc} as well as recent 
measurements of the muon anomalous magnetic moment $a_\mu$ at Fermilab~\cite{Muong-2:2021ojo}. This measurement confirmed the earlier 
tension observed at BNL~\cite{Muong-2:2006rrc} from the standard model (SM) expectation~\cite{Aoyama:2020ynm}.
Sources of uncertainty that affect the SM prediction have given further motivation to constraining the hadronic polarisation contribution to $\Delta a_\mu^{\text{had}}$ 
as there exists significant deviation between some lattice QCD predictions~\cite{Borsanyi:2020mff,Ce:2022kxy,Alexandrou:2022amy}
and analytically driven techniques based on the $R$ ratio~\cite{Keshavarzi:2018mgv}. 

The MUonE experiment~\cite{Abbiendi:2016xup} proposed at CERN aims to achieve a measurement of  $\Delta a_\mu^{\text{had}}$  with a statistical uncertainty of around 0.3\% through a precision measurement of $\mu-e$ scattering of a 150~GeV muon beam off atomic electrons, collecting $\sim 3.7\times 10^{12}$ scattering events resulting from a (leading order) cross section of around $250~\mu \text{b}$. Such scattering at
relatively low centre-of-mass energy implies a small $t$ channel momentum exchange compared to the $Z$ boson threshold thus enabling the direct measurement of $\Delta a_\mu^{\text{had}}$ (see also~\cite{Keshavarzi:2018mgv,Balzani:2021del}). 
The expected experimental precision is reflected by the continued effort to reduce the theoretical uncertainty of the SM prediction of $\mu-e$ scattering~\cite{Mastrolia:2017pfy,Gakh:2018sat,DiVita:2018nnh,CarloniCalame:2020yoz,Banerjee:2021mty,Budassi:2021twh}; in parallel such precise measurements have implications for the BSM physics as well, both in relation to $a_\mu$~\cite{Dev:2020drf} and beyond~\cite{Masiero:2020vxk,Asai:2021wzx,Galon:2022xcl}. Modifications due to new electroweak states could be constrainable at the MUonE environment, in particular when they appear at relatively low masses away from the decoupling limit, where they might be difficult to observe at hadron colliders due to overwhelming QCD backgrounds. 

Particularly interesting in this context is the two Higgs doublet model, which continues to have appealing phenomenological properties with regards to the UV completion of the SM~\cite{Dorsch:2014qja,Dorsch:2016tab,Basler:2016obg,Basler:2018dac,Basler:2019nas,Goncalves:2021egx,Su:2020pjw,Wang:2021ayg,Dorsch:2017nza,Atkinson:2021eox,Basler:2021kgq,Atkinson:2022pcn,Anisha:2022hgv}. For $\mathbb{Z}_2$ symmetry assignments of the two Higgs doublet models (2HDMs) that avoid tree-level neutral current flavour violation~\cite{Glashow:1976nt}, we will show that there are parameter regions that can explain the $a_\mu$ anomaly whilst avoiding Large Electron Positron (LEP), LHC and flavour constraints (see also~\cite{Athron:2021iuf,Atkinson:2021eox,Atkinson:2022pcn}).

Any significant sensitivity to new physics beyond the hadronic vacuum polarisation will require due caution when MUonE results are interpreted along the latter lines. The question of the extent to which a precision  experiment like MUonE can provide additional sensitivity via modified electroweak corrections is therefore relevant and timely. This is the focus of this work: we include the MUonE sensitivity estimate from next-to-leading order electroweak corrections arising in 2HDMs to a global analysis of LHC, LEP, electroweak precision and flavour physics data. We particularly focus on the relevance of lepton-specific $\mathbb{Z}_2$ charge assignments that naively tension MUonE against flavour and collider measurements. 

In addition, very recently, the CDF collaboration reported a new measurement of the $W$-boson mass $m_W$ \cite{CDF:2022hxs} which finds an excess of $\sim70\,$MeV ($7\sigma$) above the Standard Model fit \cite{ParticleDataGroup:2020ssz}. 
If such a deviation from the Standard Model is confirmed, this can open the door for a plethora of New Physics (NP) models which could explain this anomaly. 
Previously \cite{Lopez-Val:2012uou} it was expected that any 2HDM would only ever have much smaller corrections to $m_W$ than the deviation found in \cite{CDF:2022hxs}, however revived analyses of flavour conserving 2HDMs with a softly broken ${\mathbb{Z}}_2$ symmetry claim larger contributions can arise, which could bridge this gap \cite{Bahl:2022xzi,Song:2022xts,Heo:2022dey,Ahn:2022xeq,Lee:2022gyf,Benbrik:2022dja,Abouabid:2022lpg,Botella:2022rte,Kim:2022xuo,Kim:2022hvh}. 
It is therefore worthwhile to comment, in passing, on the corrections to the $W$-boson mass within the 2HDM using the context of our fits from other sectors to inform the parameter regions used, resulting in ranges of values which $m_W$ could take from a particular 2HDM realisation.
These ranges can then be compared to future updates of the measurement of $m_W$ from any experiment.
Furthermore, we correlate these ranges with the corrections to $a_\mu$ in the 2HDM, where we assume the Theory Initiative's SM prediction~\cite{Aoyama:2020ynm}, and find 2HDM parameters which can resolve either or both of these tensions with experiment. 

This work is organised as follows: Section~\ref{sec:model} provides a brief overview of the models studied in this work to make this paper self-consistent. Section~\ref{sec:muone} gives a summary of these scenarios in the phenomenological context of MUonE. Section~\ref{sec:results} is devoted to results. We summarise and conclude in Section~\ref{sec:conc}.

\begin{table}[!b]
    \centering
    \renewcommand{\arraystretch}{1.2}
    \begin{tabular}{|c||c|c|c|c|}
        \hline
        \rm Model & I & II & X & Y \\
        \hline\hline
        $u_R$ & $\Phi_2$  & $\Phi_2$  & $\Phi_2$  & $\Phi_2$ \\
        \hline
        $d_R$ & $\Phi_2$  & $\Phi_1$  & $\Phi_2$  & $\Phi_1$ \\
        \hline
        $e_R$ & $\Phi_2$  & $\Phi_1$  & $\Phi_1$  & $\Phi_2$ \\
        \hline
    \end{tabular}
    \caption{The four Types of 2HDM that avoid tree level FCNC. By convention the up-type quarks couple to $\Phi_2$.} 
    \label{tab:types}
\end{table}

\section{Models}
\label{sec:model}

The 2HDM invokes a pair of distinct complex $SU_L(2)$ doublets, $\Phi_{1,2}$, where the SM has only a single doublet that serves to give mass terms to both the up and down-type quarks. The vacuum expectation values (VEVs) of these doublets are non-zero and satisfy $v_1^2 +v_2^2 = v^2$ for the SM VEV $v \approx 246$~GeV. The general potential for this 2HDM is, in the notation of~\cite{Branco:2011iw,Gunion:1989we},
\begin{multline}
\label{eq:potential}
    V(\Phi_1,\Phi_2)=m_{11}^2\Phi_1^\dagger\Phi_1+m_{22}^2\Phi_2^\dagger\Phi_2
    -m_{12}^2(\Phi_1^\dagger\Phi_2+\Phi_2^\dagger\Phi_1)\\
    + \frac{\lambda_1}{2}(\Phi_1^\dagger\Phi_1)^2+\frac{\lambda_2}{2}(\Phi_2^\dagger\Phi_2)^2 +
    \lambda_3 (\Phi^ \dagger_1\Phi_1) (\Phi^\dagger_2\Phi_2) \\
    + \lambda_4 (\Phi^\dagger_1\Phi_2) (\Phi^\dagger_2\Phi_1) +
    \frac{\lambda_5}{2} \left[(\Phi^\dagger_1\Phi_2)^2+(\Phi^\dagger_2\Phi_1)^2\right].
\end{multline}
This opens up the possibility of tree level flavour changing neutral currents (FCNC), which can be removed by imposing a $\mathbb{Z}_2$ symmetry such that each fermion type couples to only one of the Higgs doublets. There are four such configurations, shown in Table~\ref{tab:types}, and thus four Types of 2HDM which we examine here; we denote these by, for instance, 2HDM-I.

Electroweak symmetry breaking sees three of the eight degrees of freedom of the doublets ``eaten" by the weak bosons, the five remaining degrees of freedom manifest as new particles; two neutral scalars $h^0$, $H^0$ ($m_{h^0} < m_{H^0}$), two charged scalars, $H^\pm$ and a neutral pseudoscalar~$A^0$. Instead of the lambda basis of Eq.~(\ref{eq:potential}), in this work we make use of the mass basis, which can be attained through the transformations given in \cite{Atkinson:2021eox,Basler:2016obg,Arnan:2017lxi,Han:2020zqg,Kling:2016opi}, which allows us to focus on the physical masses $m_{h^0}$, $m_{H^0}$, $m_{H^\pm}$, $m_{A^0}$, the mixing angles $\cos(\beta-\alpha)$, $\tan\beta = v_2/v_1$ and the softly $\mathbb{Z}_2$ breaking parameter $m_{12}^2$. We additionally take $h^0$ to be the scalar particle observed at the LHC, thereby fixing $m_{h^0} = 125.25\pm0.17$~GeV~\cite{ParticleDataGroup:2020ssz}.

Having transformed to the mass basis, the Yukawa sector of the Lagrangian of the general 2HDM is given by~\cite{Crivellin:2019dun, Branco:2011iw}
\begin{align}
	\nonumber
      {\cal L}^{\rm 2HDM}_{\rm Yukawa} & 
      =  -  \!\! \sum \limits_{f= u,d,\ell} \! \! \frac{m_f}{v} 
    \left(\xi_h^f \, \bar{f} f h + \xi_H^ f \, \bar{f} f H + i \eta_f \xi_A^f  \, \bar{f} \gamma_5 f A \right) \\
       \nonumber & + \left[
        \frac{\sqrt{2} V_{ud}}{v} \bar{u} 
        \left(m_d \, \xi_A^d P_R - m_u \, \xi_A^u P_L \right) d H^+ \right. \\
        & \left. +  \frac{\sqrt{2}}{v} m_\ell \, \xi_A^l  (\bar{\nu}  P_R  \ell)  H^+ + {\rm h.c.}
        \right] \! ,
        \label{eq:yukawa}
\end{align}
where we follow the convention of \cite{Crivellin:2019dun} for the $H^+$ coupling. The couplings $\xi$ are Type dependent and given in Table~\ref{tab:xis}, and the factor $\eta_f$ for fermion type $f=d,\ell$ is $1$ and for $f=u$ is $-1$.

In previous works \cite{Atkinson:2021eox, Atkinson:2022pcn} we have studied the 2HDM-I and 2HDM-II in depth and here bring the same machinery to bear on the 2HDM-X and 2HDM-Y (also known as lepton-specific and flipped respectively), as well as investigating the implications from the high precision MUonE experiment for all four models.

\begin{table}[!t]
    \centering
    \renewcommand{\arraystretch}{1.2}
    \begin{tabular}{|c||c|c|c|c|}
        \hline
        \rm Model & I & II & X & Y \\
        \hline\hline
        $\xi^u_h$ & $\cos\alpha/\sin\beta$ & $\cos\alpha/\sin\beta$ & $\cos\alpha/\sin\beta$ & $\cos\alpha/\sin\beta$ \\
        \hline
        $\xi^d_h$ & $\cos\alpha/\sin\beta$ & $-\sin\alpha/\cos\beta$ & $\cos\alpha/\sin\beta$ & $-\sin\alpha/\cos\beta$ \\
        \hline
        $\xi^l_h$ & $\cos\alpha/\sin\beta$ & $-\sin\alpha/\cos\beta$ & $-\sin\alpha/\cos\beta$ & $\cos\alpha/\sin\beta$ \\
        \hline\hline
        $\xi^u_H$ & $\sin\alpha/\sin\beta$ & $\sin\alpha/\sin\beta$ & $\sin\alpha/\sin\beta$ & $\sin\alpha/\sin\beta$ \\
        \hline
        $\xi^d_H$ & $\sin\alpha/\sin\beta$ & $\cos\alpha/\cos\beta$ & $\sin\alpha/\sin\beta$ & $\cos\alpha/\cos\beta$ \\
        \hline
        $\xi^l_H$ & $\sin\alpha/\sin\beta$ & $\cos\alpha/\cos\beta$ & $\cos\alpha/\cos\beta$ & $\sin\alpha/\sin\beta$ \\
        \hline\hline
        $\xi^u_A$ & $\cot\beta$ & $\phantom{-}\cot\beta$ & $\phantom{-}\cot\beta$ & $\phantom{-}\cot\beta$ \\
        \hline
        $\xi^d_A$ & $\cot\beta$ & $-\tan\beta$ & $\phantom{-}\cot\beta$ & $-\tan\beta$ \\
        \hline
        $\xi^l_A$ & $\cot\beta$ & $-\tan\beta$ & $-\tan\beta$ & $\phantom{-}\cot\beta$ \\
        \hline
    \end{tabular}
    \caption{Coupling strengths $\xi$ in each Type of 2HDM between the Higgs particles and fermions.}
    \label{tab:xis}
\end{table}

\section{2HDMs at MUonE}
\label{sec:muone}
\begin{figure*}[!t]
\includegraphics[width=0.9\textwidth]{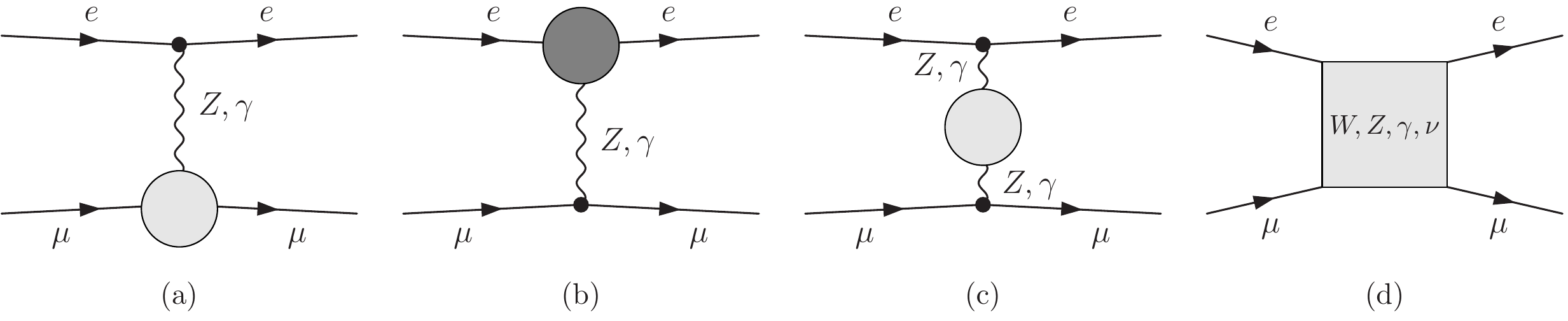}
\caption{Feynman diagram topologies contributing to $e-\mu$ scattering at one loop level. Highlighted are renormalised vertex and propagator corrections for the topologies (a)-(c). They include 2HDM Higgs contributions for the propagator and $\mu$ vertex corrections. The different shading of (b) indicates that we are neglecting electron mass contributions to the virtual amplitude. The propagator contributions include leading order hadron polarisation effects consistently.
\label{fig:feyndiags}}
\end{figure*}
The MUonE experiment aims to investigate elastic $e-\mu$ scattering. The Born-level $e-\mu$ scattering is entirely SM-like, which highlights the precision potential of the MUonE experiment to access virtual contributions that are modified from the SM expectation. At (renormalised) one-loop level the virtual ${\cal{M}}_{\text{virt}}$ part of the next-to-leading order amplitude
\begin{equation}
|{\cal{M}}|^2_1=|{\cal{M}}_\text{Born}|^2 + 2 \text{Re}\left\{ {\cal{M}}_\text{Born} {\cal{M}}^\ast_\text{virt} \right\} 
\end{equation}
can be diagrammatically represented by the Feynman diagram toplogies of Fig.~\ref{fig:feyndiags}. As we are interested in the scenarios of Table~\ref{tab:xis} where the couplings are lepton-flavour universal, and as $m_e\ll m_\mu$, we will neglect effects $\sim m_e$ contributing to the virtual amplitude throughout this work. This removes $t$-channel Higgs and Goldstone diagram contributions as well as scalar contributions to the box diagrams in Fig.~\ref{fig:feyndiags}(d); we employ the on-shell renormalisation scheme in the following. It is worthwhile highlighting that MUonE probes a subset of oblique corrections that are included in our scan as detailed in~Section~\ref{sec:results}, however, probed at space-like $t$-channel momentum transfers (the latter also means that width contributions to the $Z$ boson propagators are absent).

While the $t$-channel photon contribution already results in a soft singularity at tree-level (which is avoided by requiring a finite recoil energy in the experimental setup), the presence of virtual massless propagators connecting on-shell legs leads to additional soft (eikonal) singularities of $|{\cal{M}}|^2_1$ which cancel against real photon emission via the Kinoshita-Lee-Nauenberg theorem~\cite{Kinoshita:1962ur,Lee:1964is}. For abelian gauge theories the eikonal approximation takes a particularly compact form that is well-documented in the literature~\cite{tHooft:1978jhc,Denner:1991kt} and is readily implemented in publicly available packages like {\sc{FormCalc}}~\cite{Hahn:1998yk}. This real emission part is not sensitive to the 2HDM modifications, but we consider it for completeness, including unresolved soft photon radiation up to 10\% of the MUonE centre-of-mass energy.

Numerical expressions of the loop diagrams sketched in Fig.~\ref{fig:feyndiags} are generated using the {\sc{FeynArts}}/{\sc{FormCalc}}/{\sc{LoopTools}}~\cite{Hahn:1999mt,Hahn:1998yk,Hahn:2001rv} toolchain. We have checked the cancellation of ultraviolet divergences both numerically and analytically for the full amplitude, as well as the independence of our results of a virtual photon mass that is introduced to regularise soft singularities at intermediate steps. We have also validated our results against the SM by approaching the 2HDM decoupling limits of the different scenarios considered in this work.

As a $2\to 2$ scattering process, MUonE is entirely determined by the $t$ channel momentum transfer (we assume mono-chromatic muon beams and electron targets), which determines the laboratory scattering angles and $\mu$ and $e$ recoil energies. For our analysis we consider 51 independent bins of a scattering angle distribution (after validating the literature-documented cross section of $\sim 240~\mu\text{b}$), which we enter in a $\chi^2$ hypothesis test given~by
\begin{equation}
	\chi^2 = \sum_i {(N_i - N_i^\text{SM})^2\over \sigma^2_{i,\text{syst}} + \sigma^2_{i,\text{stat}}}\,.
\end{equation}
Here, $i$ runs over the bins, $N_i$ are the 2HDM events in the $i$th bin and $N_i^\text{SM}$ denotes the SM expectation (evaluated at next-to-leading order). $\sigma_{i,\text{stat}}, \sigma_{i,\text{syst}}$ denote statistical and systematic uncertainties, respectively. In the following we will use $\sigma_{i,\text{stat}}=\sqrt{N_i}$ and $\sigma_{i,\text{syst}}=10^{-5} N_i$ as a common benchmark choice for the expected sensitivity of MUonE~\cite{Abbiendi:2016xup,Dev:2020drf}. From the $\chi^2$ we can infer the impact of a given parameter point on the MUonE data and whether the experiment would be sensitive to the existence of the 2HDM with such parameters.

\section{Constraints, Results, and MUonE relevance}
\label{sec:results}
Here we outline the various constraints that can be placed on the 2HDM from a variety of sectors. We begin with an analysis of the signal strengths of the SM-like Higgs observed at the LHC, proceeding to flavour observables where the 2HDM can have large indirect effects, and then to the impact of direct BSM searches on the parameter space, including an extrapolation to future collider capabilities for the latter. These analyses build on the toolchains developed in our previous works~\cite{Atkinson:2021eox,Atkinson:2022pcn}; we make use of the python package \texttt{flavio}~\cite{Straub:2018kue} for the flavour fits and \texttt{MadGraph5\_aMC@NLO}~\cite{Alwall:2014hca},  \texttt{2HDecay} \cite{Krause:2018wmo, Djouadi:1997yw, Djouadi:2018xqq, Krause:2016xku, Denner:2018opp, Hahn:1998yk}, and \texttt{HiggsBounds} \cite{Bechtle:2008jh, Bechtle:2011sb, Bechtle:2012lvg, Bechtle:2013wla, Bechtle:2015pma, Bechtle:2020pkv, Bahl:2021yhk} for the BSM collider searches. We refer the interested reader to~\cite{Atkinson:2021eox,Atkinson:2022pcn}  for full details on the fitting procedures used and experimental data included, updated to additionally include the measurements from~\cite{CMS:2022psv, ATLAS:2022tnm}, where we will avoid being overly repetitive here and focus on the conclusions drawn from these analyses.

\begin{figure*}[!t]
	\begin{center}
		\includegraphics[width=0.48\textwidth]{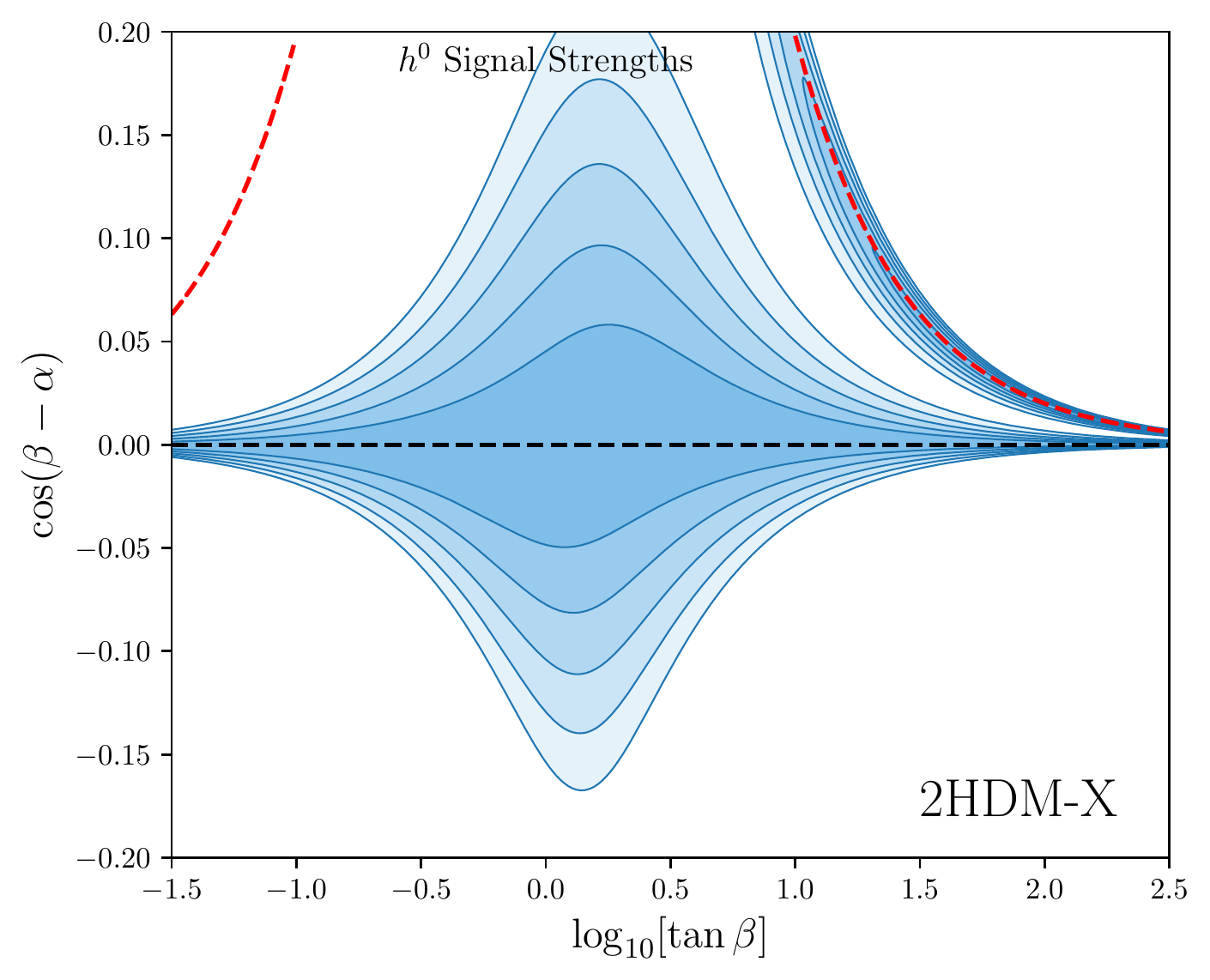}
    		\includegraphics[width=0.48\textwidth]{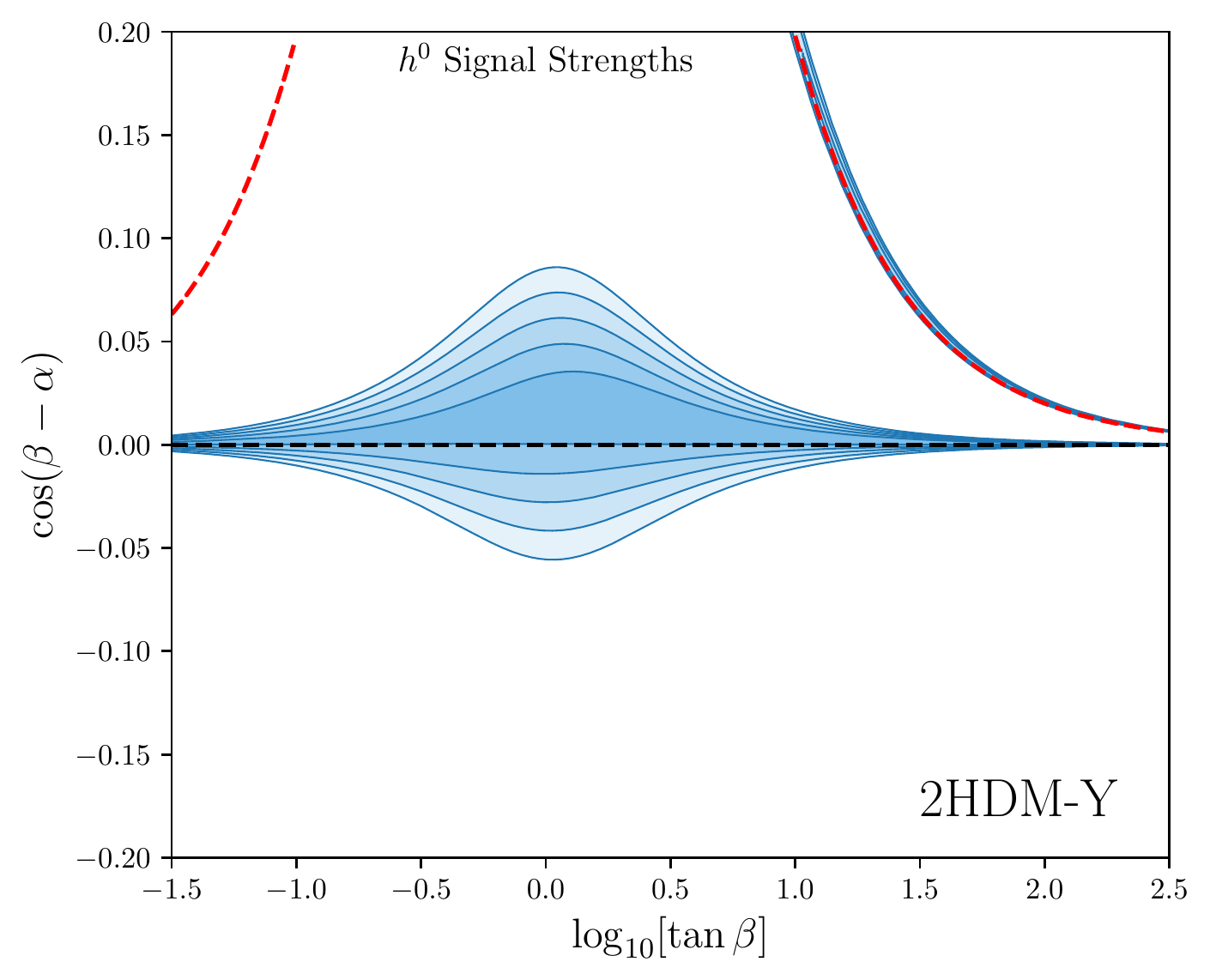}
		\caption{\label{fig:signals} Contour plots of the allowed regions at 1, 2, 3, 4, $5\sigma$ from the $h^0$ signal strengths with the alignment and wrong sign limits shown as black and red dashed lines respectively}
	\end{center}
\end{figure*}

\subsection{Higgs Signal Strengths}
\label{sec:signals}
The signal strengths of the observed Higgs boson provide a measure of how closely this boson matches the expected phenomenology of the SM Higgs, and are defined as, for a channel with production mechanism $i$ and decay products $f$,
\begin{equation}
\mu^f_i = \frac{(\sigma_i \cdotp {\cal B}_f)_\text{Exp.}}{(\sigma_i \cdotp {\cal B}_f)_\text{SM}}\,.
\label{eq:sigs_def}
\end{equation}
The modifications to the couplings of $h^0$ given in Table~\ref{tab:xis} lead to different phenomenology in the 2HDM compared to the SM, and we use this to constrain the parameters on which the couplings depend; $\tan\beta$ and $\cos(\beta-\alpha)$ in the basis we use here. As the observed Higgs matches the SM predictions to within $\sim 10$\%, we expect to find that the 2HDM is driven towards the alignment limit, $\cos(\beta-\alpha)=0$, which precisely recovers the SM couplings. There is another region in which SM-like couplings can be recovered, with, in the 2HDM-X, $\xi^u=\xi^d=1$ and $\xi^\ell=-1$. This is known as the wrong sign limit and is attained when
\begin{equation}
    \label{eq:WSL_cond}
    \cos{(\beta - \alpha)} = \sin{2\beta} = \frac{2 \tan \beta}{1 + \tan^2 \beta} \, .
\end{equation}
Similar coupling configurations in which the $\xi^i$ have the same magnitude as the SM are also possible in Type II and Y models, with the down-type and lepton and only the down-type couplings taking the negative sign respectively. The signal strengths are calculated using the expressions in~\cite{Gunion:1989we}, with the results of the fit shown in Fig.~\ref{fig:signals}.

In the 2HDM-X we find that the wrong sign limit is allowed for all confidence levels examined here. Outside this region, there is a maximum magnitude of $|\cos(\beta-\alpha)| \leq 0.10$ at 2$\sigma$, which falls rapidly once $\tan\beta$ is away from $\sim$ 1, with only very small $\cos(\beta-\alpha)$ allowed for extreme values of $\tan\beta$. The wrong sign limit being allowed at 1$\sigma$, in contrast to the 2HDM-II in which it is excluded up to 2.7$\sigma$~\cite{Atkinson:2021eox}, is a result of the lack of sensitivity to the sign of $\xi^\ell$, due to the leptons giving minimal contributions to the loop level processes compared to the much more important contributions from the quarks. Indeed, the leptons do not contribute at all to the crucial gluon fusion production mechanism and their contributions to the diphoton and $Z\gamma$ decay channels are orders of magnitude lower than those from the quarks and the $W^\pm$ boson for an SM-like $h^0$. The leptonic decays of $h^0$ are still important here as they restrict $\cos(\beta-\alpha)$ to be small for large values of $\tan\beta$, which differentiates these results from the 2HDM-I. 

The relative lack of sensitivity to $\xi^\ell$ compared to the quark couplings can also be seen in the results for the 2HDM-Y, in which the quark sector matches that of the 2HDM-II, with the result that the contours are very similar between the two Types, with the wrong sign limit allowed at 2.6$\sigma$ in the 2HDM-Y, with  $|\cos(\beta-\alpha)| \leq 0.049$ at 2$\sigma$.

\subsection{Collider Searches, Electroweak Precision and Flavour Constraints}
\label{sec:collider}
\begin{figure*}[tbh]
	\begin{center}
		\includegraphics[width=0.48\textwidth]{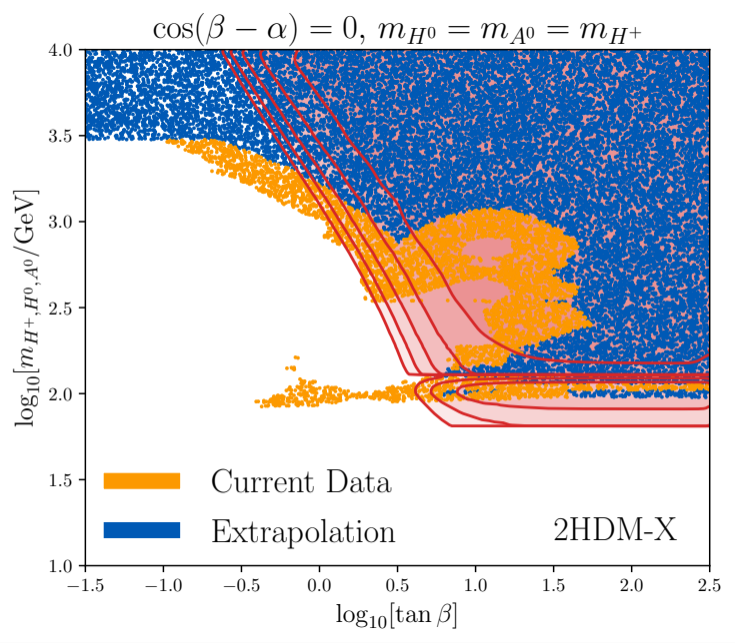}
    		\includegraphics[width=0.48\textwidth]{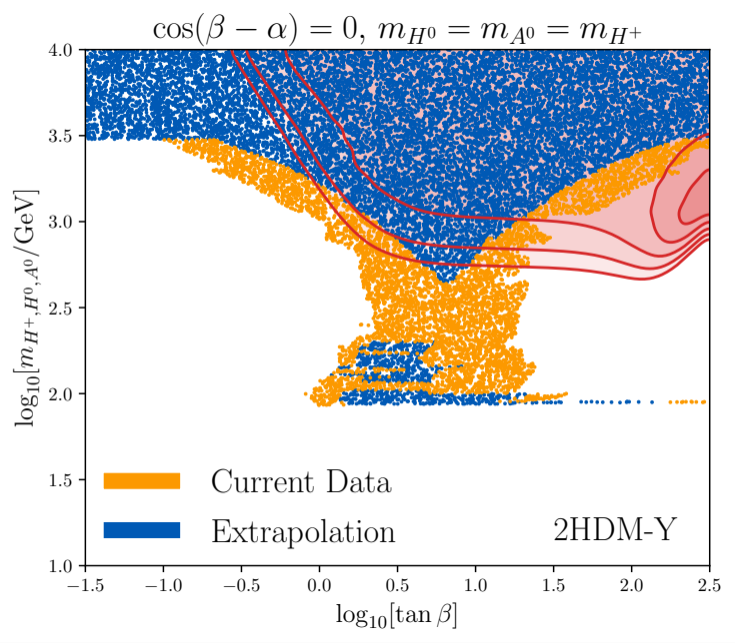}
		\caption{\label{fig:collider} Scans of the 2HDM parameter space with randomly generated points shown in blue if allowed by the current and extrapolated bounds, and in orange if currently allowed but expected to be excluded by the HL-LHC. Overlaying are the 1, 2, 3, 4, 5$\sigma$ (darker to lighter) allowed contours from the flavour sector to highlight regions of collider-flavour (including electroweak precision data) complementarity.}
	\end{center}
\end{figure*}

In this section we combine the numerous collider searches for BSM Higgs states to exclude regions of the parameter space. We take the exact alignment and degenerate mass limits in these scans, guided by the results of Section~\ref{sec:signals} and the theoretical constraints~\cite{Atkinson:2021eox}. We also extrapolate the LHC data present in \texttt{HiggsBounds} to a future collider with a centre of mass energy of $\sqrt{s}=13$~TeV and $\mathcal{L} =3\,\text{ab}^{-1}$, in line with the expected integrated luminosity of the HL-LHC. The results from a scan of 50,000 randomly generated points are shown in Fig.~\ref{fig:collider}, where orange points are allowed by current data but excluded by the extrapolated dataset while blue points are allowed by both. In the following we outline the main channels that give the exclusion in each region. 

At low $\tan\beta$ in the 2HDM-X, the exclusion is from decays of the neutral heavy Higgses to $b$ quarks~\cite{ALEPH:2006tnd,CMS:2018pwl}, up to the top quark mass, beyond which $H^+ \to t\bar{b}$ drives the exclusion until the mass of the charged Higgs becomes large enough such that the production cross section falls below the current bounds~\cite{ATLAS:2020jqj}. For moderate $\tan\beta$ the LEP baseline and $H^+ \to \tau^+\bar{\nu_\tau}$ rule out masses up to $\sim$ 100 GeV~\cite{CMS:2018ect} before sensitivity is lost due to falling branching ratios for this channel. The decay $H^0 \to \tau^+\tau^-$ is responsible for the remaining exclusion in this region~\cite{ATLAS:2020zms}. Whilst we might expect this channel to also exclude points with large $\tan\beta$ as the lepton couplings become large, this is not the case as the quark couplings, proportional to $\cot\beta$, become small and the new Higgs states essentially decouple from the quark sector, giving very low production cross sections and thus limited sensitivity in this region beyond the historic LEP limits~\cite{ALEPH:2006tnd}. The extrapolation improves these bounds, leading to the additional exclusion most notable in the moderate $\tan\beta$ region from $H^0 \to \tau^+\tau^-$.

In the 2HDM-Y case, below the top mass at low $\tan\beta$, leptonic decays of $H^0$ exclude points~\cite{CMS:2019lwf,CMS:2015mca}, before the same $H^\pm \to t\bar{b}$ channel provides the exclusion above the top mass~\cite{ATLAS:2020jqj}. There is relatively little exclusion for moderate $\tan\beta$ in the current data, owing to the comparatively low branching ratios in this region for the channels that give exclusion in other 2HDM Types, with the lower mass bound set from the LEP data~\cite{ALEPH:2006tnd}. For high $\tan\beta$, $H^0 \to b\bar{b}$ gives the exclusion up to $\sim$ 250 GeV~\cite{CMS:2015grx}, beyond which $H^+ \to t\bar{b}$ is the most sensitive channel. In the extrapolation, the improved bounds from $H^+ \to t\bar{b}$ and $H^0 \to b\bar{b}$ are sufficient to rule out a swathe of points in the moderate $\tan\beta$ region. 

Furthermore, we find constraints on the 2HDM parameters by way of a global fit to flavour and electroweak precision observables, where the
$h^0$ signal strengths are also included. 
These fits for Types I and II have been previously covered in \cite{Atkinson:2021eox,Atkinson:2022pcn}, and we also now consider Types X and Y. 
Having updated several input parameters, the predictions for many of the flavour observables have changed from the previous listings in \cite{Atkinson:2021eox}. Although not significant enough to have a noticeable impact on these fits, we collect the SM predictions for the observables used in an auxiliary file. 
Note that here by default we assume the global fit including all observables {\it apart from} the LFU quantities $R(D^{(*)}), R(K^{(*)})$. The latter (as it was shown in~\cite{Atkinson:2021eox,Atkinson:2022pcn})
cannot be accommodated within 2HDMs of Type II or Type I,
and we find the same for Types X and Y in this work.
Including $R(D^{(*)}), R(K^{(*)})$ in the fit therefore leads to very poor $p$-values. On the other side, 
excluding LFU observables from the fit yields more reasonable $p$-values, and therefore, in such scenarios, constraints on 2HDM parameter spaces have higher credibility.

The global fit for Type X results in very similar constraints to that of Type I, where overall there is still a lot of freedom for the 2HDM parameters.
The mass of $H^\pm$ can be as low as the electroweak scale, which is anyway the lower limit of validity for the Wilson coefficient expressions used from \cite{Crivellin:2019dun}.
Above this minimum value, the correlation between the lower bound on $\tan\beta$ and $m_{H^\pm}$ is again very similar to that of the Type I fit; we are not sensitive to the upper limits on $\tan\beta$ as for large values, we approach decoupling in the quark sector and non-perturbative coupling for the leptons.

The charged Higgs mass is most constrained in the Type Y model, where we find strict lower and upper limits within $2\sigma$, shown in Table~\ref{tab:bounds}.
Beyond $2\sigma$, we find mass constraints more similar to those of Type II where these two Types have the same quark sector couplings, leading to agreement in the strong constraint on $m_{H^\pm}$ from ${\cal B}(\bar{B}\to X_s\gamma)$. 

We now look to compare the results of the BSM collider searches and the global fits to gain insight into the complementary nature of these approaches. We do so by displaying the results on the same axes in Fig.~\ref{fig:collider}, and stress that this is not a statistical combination and simply an overlay of the contours from the flavour sector on the scans from the collider searches in Fig.~\ref{fig:collider}. 

We observe a good degree of complementarity between the two sectors, as each can probe regions of the parameter space that the other cannot. For both Types of 2HDM examined here the flavour sector excludes regions with low $\tan\beta$ that the BSM searches lack sensitivity to, both in the current and extrapolated datasets. For the 2HDM-X, the collider searches, particularly the $H^0 \to \tau^+\tau^-$ channel in the HL-LHC extrapolation, are more sensitive than the flavour sector and can rule out a portion of the 1$\sigma$ region from the flavour observables, though both approaches lack sensitivity above masses of 100 GeV when $\tan\beta$ is large. In the 2HDM-Y case, the collider searches rule out the entirety of the 1$\sigma$ region in the parameter space we examine here, and the majority of the 2$\sigma$ region. The extrapolation of the LHC data improves on this further, and rules out very nearly all of the 2$\sigma$ region. Outside of this high $\tan\beta$ region the flavour constraints outperform the collider searches. These results demonstrate the high degree of complementarity between the two datasets and give further indications of where future searches should focus efforts to detect or exclude a 2HDM. 

\subsection{$a_\mu$ in 2HDMs}
\label{sec:2HDM_amu}
\begin{figure*}[th] 
\centering 
\includegraphics[width=0.49\textwidth]{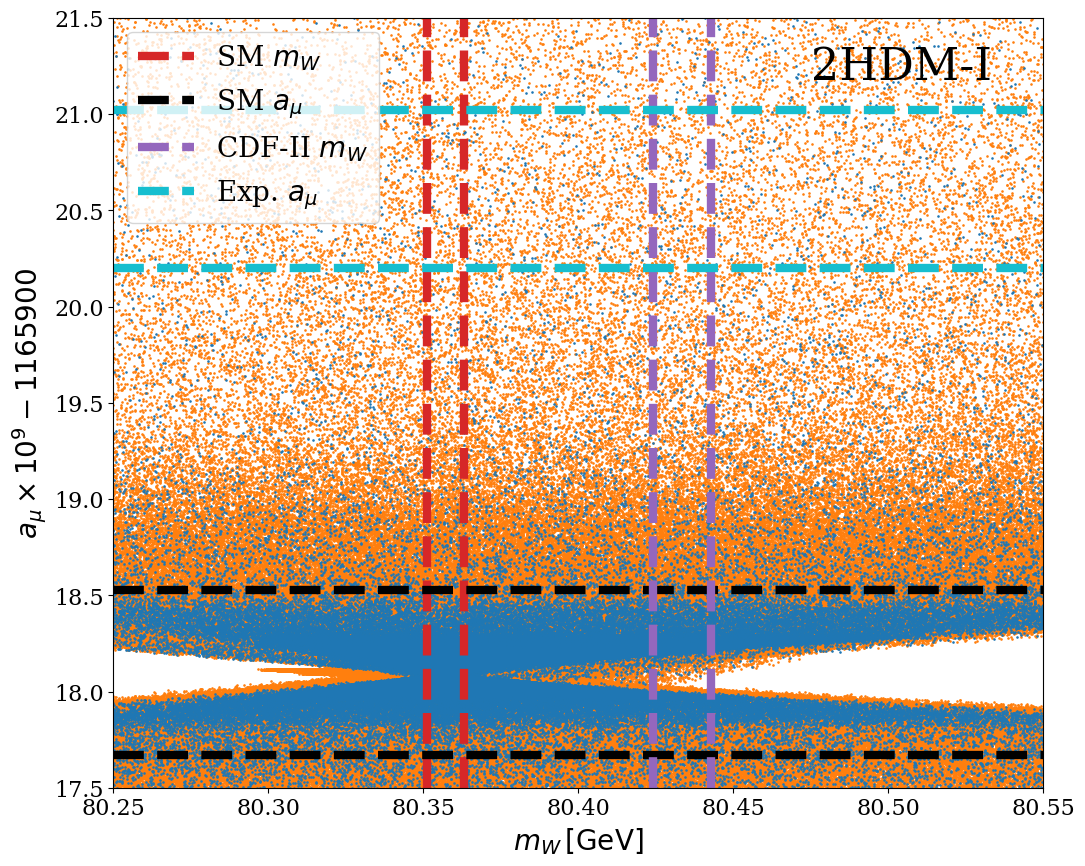} 
\includegraphics[width=0.49\textwidth]{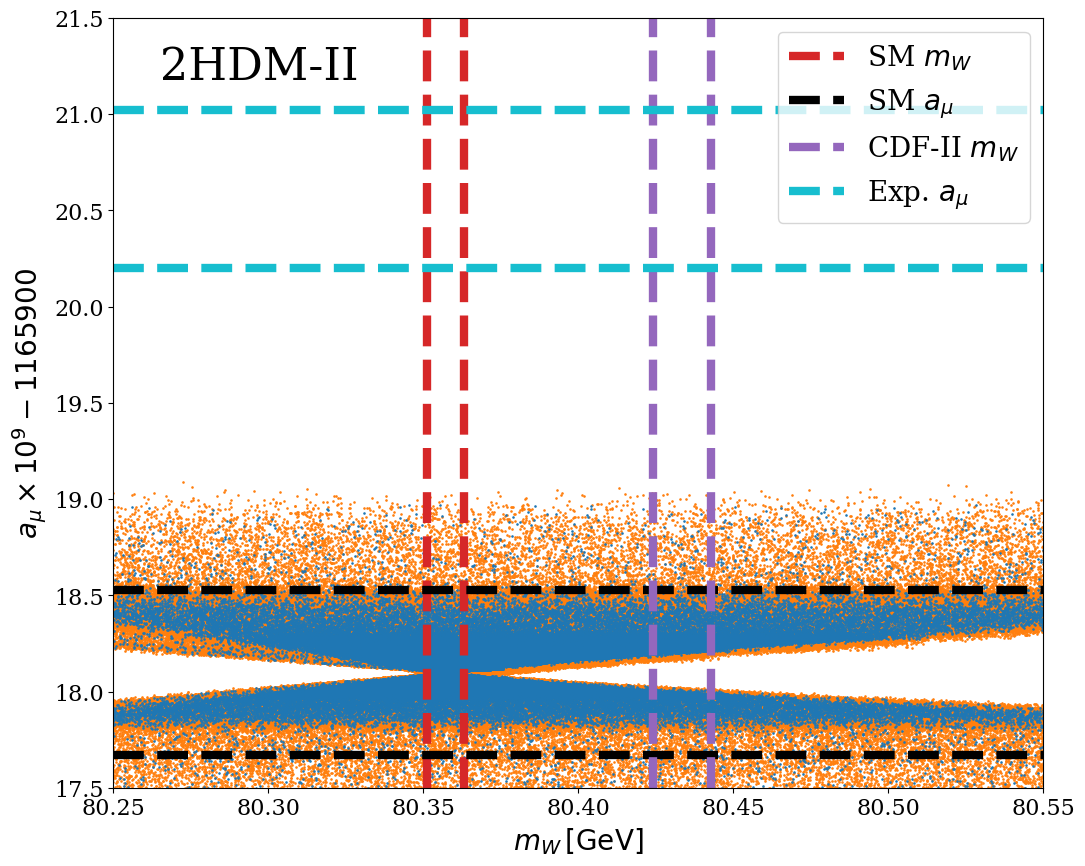}
\includegraphics[width=0.49\textwidth]{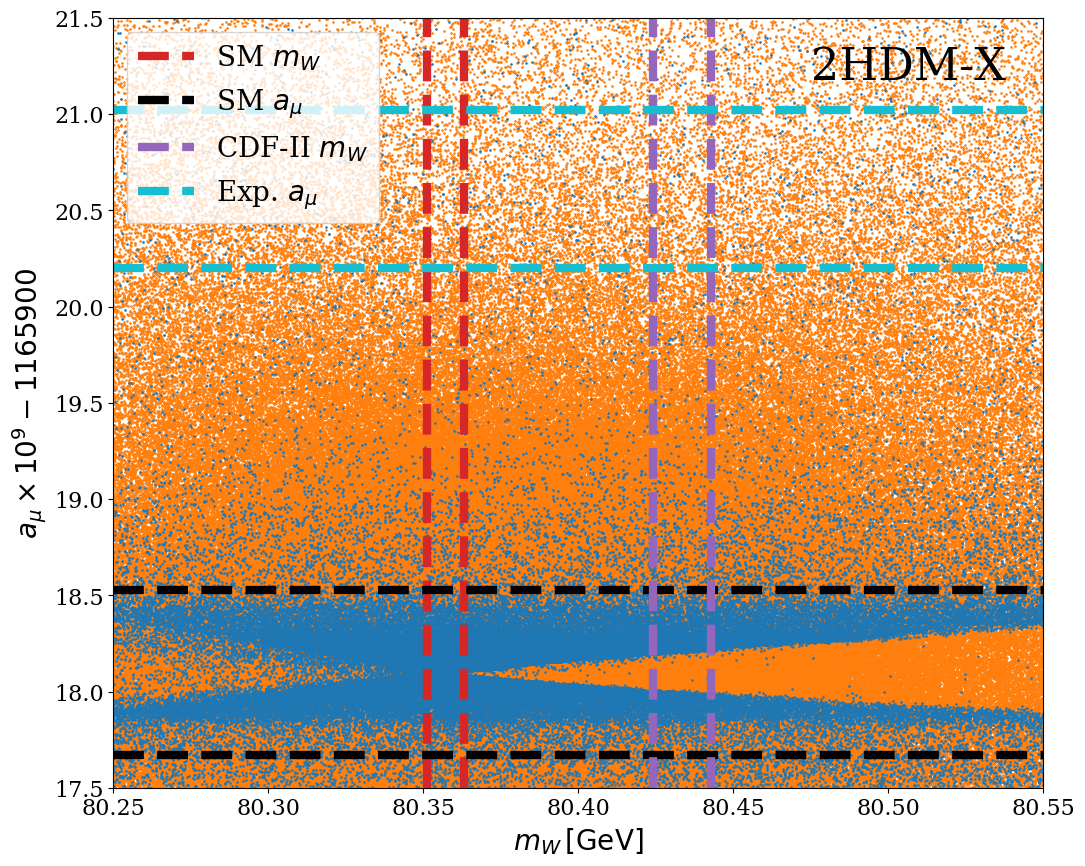}
\includegraphics[width=0.49\textwidth]{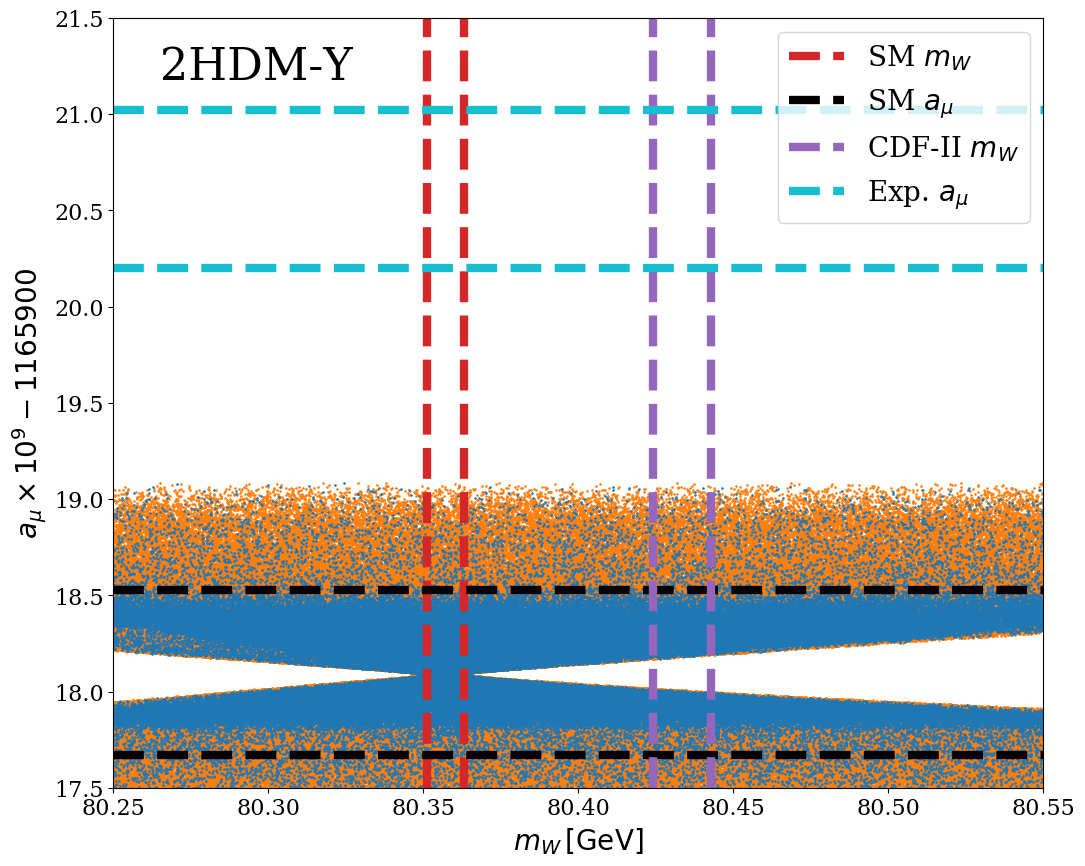}
\caption{\label{fig:resultsMB_cT0} The correlation of $m_W$ and the muon $g-2$ in 2HDMs within their allowed parameter space, see Table~\ref{tab:bounds} (orange); restricting $\cos(\beta-\alpha)=0$ (blue). The $1\sigma$ region for each quantity stated in the legend are shown by the dashed lines.}
\end{figure*}

With the long-standing tension between the SM prediction \cite{Aoyama:2020ynm} and the experimental measurement from BNL~\cite{Muong-2:2006rrc}, now also confirmed by FNAL~\cite{Muong-2:2021ojo}, there has been strong motivation to consider BSM contributions to $a_\mu$ where one can introduce new loop diagrams, such as from the new Higgs particles we consider here in the 2HDMs.
In the 2HDM, it is important to consider diagrams at both 1- and 2-loop, where the `Barr-Zee' diagrams yield significant contributions \cite{Barr:1990vd,Ilisie:2015tra}.
We will consider the 2HDM contributions to $a_\mu$ in each of the four flavour-conserving Types discussed in Section~\ref{sec:model}, where we scan over $10^7$ random points within the $2\sigma$ global-fit constraints for each Type, summarised in Table~\ref{tab:bounds}.
We also follow this procedure to analyse the 2HDM corrections to the $W$ boson mass (see Section~\ref{sec:2HDM_mW} below), and present together the correlated results for $a_\mu$ and $m_W$ in 2HDMs in Fig.~\ref{fig:resultsMB_cT0}.
Note that we assume the White Paper prediction~\cite{Aoyama:2020ynm} when referring to the SM contribution to $a_\mu$.
It is found that in order to resolve the tension in $a_\mu$ from any Type of 2HDM, one must either consider non-perturbative values of $\tan\beta$ or restrict the masses of the new Higgses to $<1\,$TeV where a sizeable mass splitting between $H^\pm,\,H^0,\,A^0$ is also allowed from theory constraints (see e.g. Table 1 of \cite{Atkinson:2021eox}).
The former scenario is one explicitly avoided in our studies and we cannot comment on the phenomenology of a non-perturbative 2HDM; the latter scenario can be compared to our fits of other sectors in 2HDMs, however, masses lower than $1\,$TeV are disfavoured at $2\sigma$ in both the Type-II and Type-Y models and thus we cannot resolve $a_\mu$ within the $2\sigma$ constraints of our global fits for Types II and Y.
In Types I and X, there is much more freedom allowed for the new Higgs masses below $1\,$TeV which leads to a much larger range of mass splittings between $H^\pm,\,H^0,\,A^0$ allowing then for larger contributions to $a_\mu$; we now find for both Types I and X many parameter points within $2\sigma$ of our global fits which would also resolve the tension in~$a_\mu$.

In Fig.~\ref{fig:resultsMB_cT0}, we plot in orange data calculated from allowing all parameters to vary within the $2\sigma$ bounds of our fits (Table~\ref{tab:bounds}), while in blue we further restrict the data to lie in the alignment limit, $\cos(\beta-\alpha)=0$. 
In comparing the blue and orange data, one can see that although deviations from alignment are expected to be small (see Section~\ref{sec:signals}), allowing these small deviations is important in resolving the tension in $a_\mu$ even in the freer Types I and X.

It is important to stress however that $a_\mu$ in the SM itself is still under scrutiny, and in fact with multiple recent lattice QCD predictions with improved precision~\cite{Borsanyi:2020mff,Ce:2022kxy,Alexandrou:2022amy} beginning to converge on a value closer to experiment than the data-driven SM prediction, the 2HDM contributions may in the future be required to be much smaller than considered here for agreement between theory and experiment to stay.

\subsection{$W$-Boson Mass}
\label{sec:2HDM_mW}
\begin{table*}[th]
    \centering
     \renewcommand{\arraystretch}{1.2}
    \begin{tabular}{|l|c|c|c|}
        \hline
        & $m_{H^+} \, {\rm [TeV]}$ & $\cos(\beta-\alpha)$ & $\tan\beta$ \\
        \hline
        Type I  & $[0.1,10]$ & $[-0.14,0.14]$ & $[0.55^{\red *},320)$ \\[3pt] 
        Type II & $[0.86,10]$ & $[-0.04,0.04]$ & $[0.32^{\red *},50^{\red *})$ \\[3pt] 
        Type X  & $[0.13,10]$ & $[-0.05,0.10]$ & $[0.38^{\red *},320)$ \\[3pt] 
        Type Y  & $[0.74,3.61]$  & $[-0.01,0.04]$ & $[110,320)$ \\
        \hline
    \end{tabular}
    \caption{\label{tab:bounds} Regions of 2HDM parameters used in this work. We consider the $2\sigma$ regions from the global fit (excluding LFU observables $R (K^{(*)}), R (D^{(*)})$) for each Type, and also do not exceed the region of $m_{H^+}\in[0.1,10]\,$TeV in order to not conflict with direct search limits \cite{ALEPH:2006tnd} or approach the decoupling limit. The additional neutral Higgs masses are constrained from the charged Higgs mass as found in \cite{Atkinson:2021eox} (Table 1). {\red *} indicates quantities which are not general, but depend on $m_{H^+}$; we show values for $m_{H^+}=10\,$TeV where the allowed regions are at their maximum.}
\end{table*}
The relation between the $W$-boson mass $m_W$ and 
the Fermi constant $G_F$, the fine-structure constant at zero momentum~$\alpha_{\rm em} (0)$,
and the $Z$-boson mass~$m_Z$ reads (e.g.~\cite{Lopez-Val:2014jva}, \cite{Hollik:2003cj}):
\begin{equation}
m_W^2 \left(1 - \frac{m_W^2}{m_Z^2} \right) 
= \frac{\pi \alpha_{\rm em} (0)} {\sqrt 2 G_F} \left(1 + \Delta r \right), 
\label{eq:mW-EW-relation1}
\end{equation}
where $m_W$ and $m_Z$ are renormalised in the on-shell scheme, and $\Delta r$ accounts for radiative corrections.
Eq.~\eqref{eq:mW-EW-relation1} can be then equivalently presented as
\begin{equation}
m_W^2 = \frac{1}{2} m_Z^2 
\left(1 + \sqrt{1- \frac{4 \pi \alpha_{\rm em} (0)} {\sqrt 2 G_F m_Z^2} \left[1 + \Delta r (m_W^2) \right]} \, \right),
\label{eq:mW-EW-relation2}
\end{equation}
which then is iteratively solved with respect to $m_W$.
To first order, 
\begin{equation}
    \Delta m_W \simeq - \frac{1}{2} m_W \frac{s_W^2}{s_W^2 - c_W^2} \Delta r,
\end{equation}
where $s_W^2 \equiv \sin^2 (\theta_W) = 1 - m_W^2/m_Z^2 $ is the shorthand notation of the electroweak mixing angle in the on-shell scheme, and $c_W^2 \equiv \cos^2 \theta_W = 1 - s_W^2$.
Introducing the oblique parameter $T$,
\begin{equation}
T \equiv \frac{1}{\alpha_{\rm em}(m_Z)} 
\left(\frac{\Sigma_Z (0)}{m_Z^2} - \frac{\Sigma_W (0)}{m_W^2}  \right),
\end{equation}
$\Delta r$ can then be recast as \cite{Lopez-Val:2014jva} 
\begin{equation}
\Delta r = \Delta \alpha - \frac{c_W^2}{s_W^2} \alpha_{\rm em} (m_Z) \, T + \Delta r_{\rm rem} \, ,
\end{equation}
where $\Delta\alpha$ includes the leading logarithmic 
QED corrections from the light-fermions,
and $\Delta r_{\rm rem}$ absorbs the remaining contributions.
For the oblique parameter $T$ we use the Yukawa-independent expression in a 2HDM shown in Eq.~(C.1) of \cite{Atkinson:2021eox}, calculated from \cite{Grimus:2008nb}.
As $T$ is independent of the Yukawa structure of the 2HDM, it is the same for each Type we discuss here; the difference between the results of the four Types will come purely through the different parameter constraints found from the global fits (Table~\ref{tab:bounds}). 

Some studies of $m_W$ in 2HDMs consider $T$ in the exact alignment limit 
\cite{Bahl:2022xzi,Botella:2022rte,Kim:2022xuo}, however here we do not constrain ourselves to this. 
Instead we take the constraints on $\cos(\beta-\alpha)$ at $2\sigma$ from the global fit to each Type and vary within these. 
The impact of choosing the exact alignment limit as opposed to our method with further freedom can be seen by comparing the blue (alignment limit) and orange ($2\sigma$ constraints) data in Fig.~\ref{fig:resultsMB_cT0}.
We do not find a significant change in the range of values $m_W$ can take in the 2HDM from the choice of treatment of $\cos(\beta-\alpha)$, instead this choice is far more important for the consideration of $a_\mu$ than
for $m_W$. In Fig.~\ref{fig:resultsMB_cT0}, both the orange and blue data are calculated using 1-loop expressions for the $T$ parameter in the 2HDM. 
There are however 2-loop calculations for the 2HDM contributions to $m_W$ calculated \cite{Hessenberger:2016atw,Hessenberger:2018xzo} and used \cite{Bahl:2022xzi} in literature, remaining in the alignment limit.
We use the package \texttt{THDM\_EWPOS}\footnote{\href{https://github.com/st3v3m4n/THDM\_EWPOS}{https://github.com/st3v3m4n/THDM\_EWPOS}} to compare the 1- and 1+2-loop calculations for $m_W$ in 2HDM. 
Within the region plotted in Fig.~\ref{fig:resultsMB_cT0}, it is found that the 1+2-loop calculation does not provide more information than the 1-loop, at least within the alignment limit, and thus we do not show additional plots for these calculations. 
The difference found at 2-loop is a `stretching out' of $m_W$ values which already extend past the plotted range at 1-loop and also that the value of $m_W$ becomes less correlated with $a_\mu$.

Within the context of our global-fit parameter constraints (at $2\sigma$), we find that the 2HDM corrections to the $W$-boson mass can be large, either positively or negatively.
To generate large corrections, it is favoured that there is a large mass splitting between $H^\pm,\,H^0,\,A^0$ which approximately implies an upper bound $m_{H^\pm}\lesssim1\,$TeV, however there is still sufficient freedom for the masses to be separate within the range of $1-2\,$TeV such that the more constrained Types II and Y can also find large corrections to $m_W$, where even the $7\sigma$ deviation of the new CDF-II measurement is easily covered by these. 

\subsection{Impact of MUonE}
\label{sec:MuoneResults}
As demonstrated in the previous section, $(g-2)_\mu$ can be accommodated in the 2HDM. Equipped with this discussion, we now turn to relevance of MUonE, which seeks to measure the hadronic vacuum polarisation contribution in a data-driven way (without relying on past measurements of the $R$ ratio). Our aim is to clarify whether new physics contributions are relevant in this context. To this end, we consider the parameter space of all four models of Tables~\ref{tab:types} and~\ref{tab:xis}. Furthermore, we consider not just the degenerate mass and alignment limits in this section, but include a wide range of possible mass scenarios, such as varying each of the new Higgs masses independently, fixing two masses to a range of values and varying the other, all with different values of $\cos(\beta-\alpha)$ examined in each case, as we scan across the mass and $\tan\beta$ parameter space.

\begin{figure}[!t]
	\begin{center}
		\includegraphics[width=0.48\textwidth]{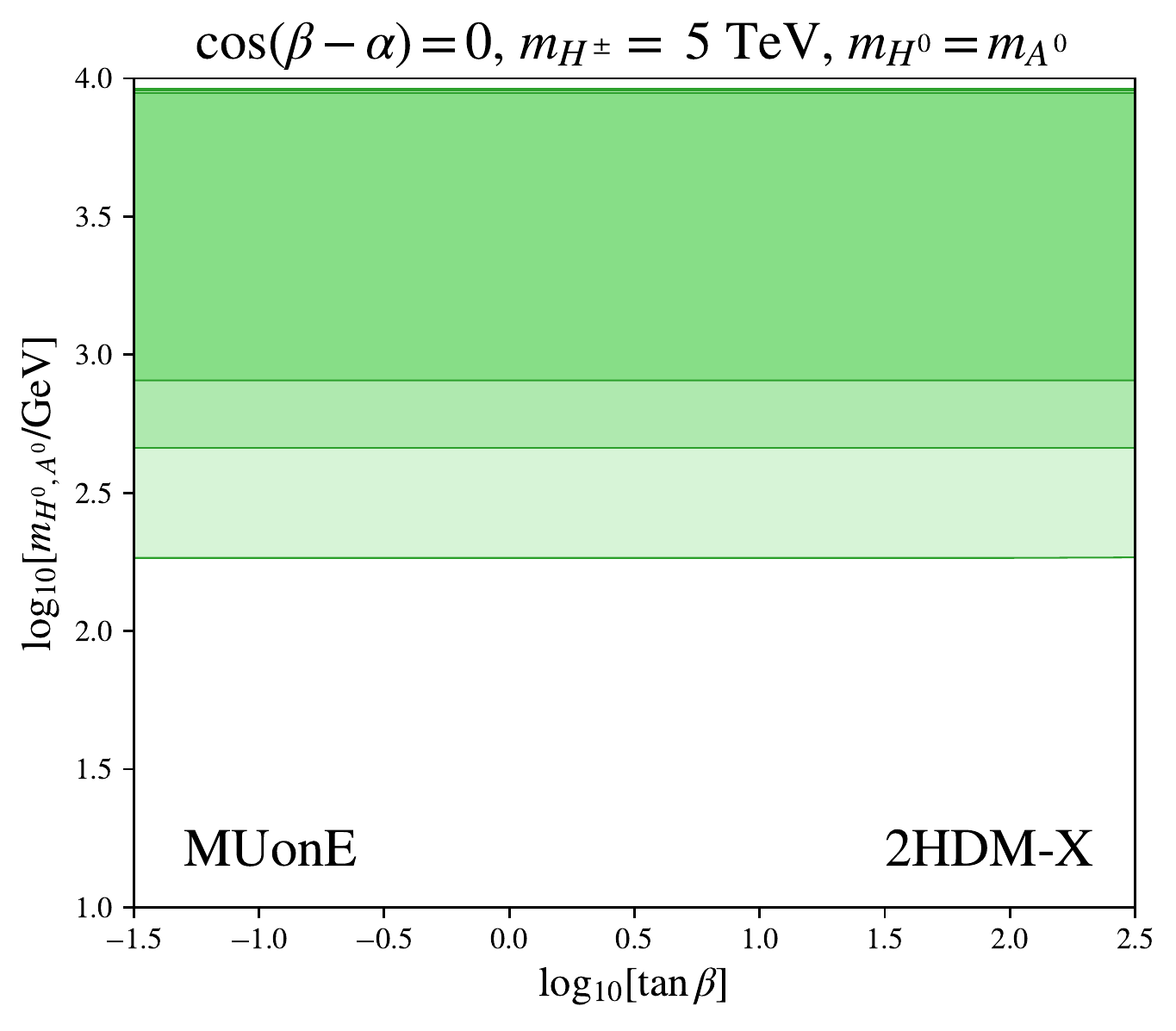}
    		\caption{\label{fig:muone} Contour plots of the allowed regions at 1, 2, 3$\sigma$ from the MUonE analysis, where in this case we have fixed $m_{H^0} = m_{A^0}$, $m_{H^\pm} =$~5 TeV and $\cos(\beta-\alpha)=0$.}
	\end{center}
\end{figure}

We find that the MUonE experiment is largely insensitive to the presence of a 2HDM of any Type. The total $\chi^2$ we calculate in almost every scenario is far smaller than the threshold value for the 68\% CL and thus we find that almost all parameter regions cannot be probed by the MUonE experiment. However, there are some exceptions to this and we can still glean some useful information from this analysis. In order to observe any exclusion we must consider extreme mass scenarios, one example of which we show for the 2HDM-X in Fig.~\ref{fig:muone}, where we have fixed $m_{H^+} = 5$~TeV, $\cos(\beta-\alpha)=0$ and varied $m_{H^0} = m_{A^0}$. In this scenario there is a $\tan\beta$ independent lower limit on the mass of the neutral scalars of $m_{H^0} = m_{A^0} \gtrsim 460$~GeV at 2$\sigma$, with an upper bound at around 8.5~TeV. Identical limits are found in this scenario for all Types of 2HDM within the parameter space we examine. Separately, there is also some exclusion at and beyond the TeV scale for $m_{H^+}=m_{H^0}$ when $m_{A^0}$ is fixed to a value below 1~TeV. Inducing a large mass splitting between the charged and neutral Higgses yields the greatest sensitivity, however such mass splits are conclusively ruled out by the theoretical requirements explored in \cite{Atkinson:2021eox}, which demand a strong degree of mass degeneracy, which becomes tighter for higher masses. 

Very low Higgs masses, below the order of 1~GeV, give large $\chi^2$, but such states would already have been observed in the direct collider searches outlined in Section~\ref{sec:collider} and thus are already ruled out, so MUonE adds no extra exclusion in this case. Additionally, points with large $\tan\beta \geq \mathcal{O}(10^4)$, are also excluded from the MUonE analysis in the 2HDM-X but such $\tan\beta$ is highly non-perturbative; a similar argument applies in the 2HDM-Y for $\tan\beta \leq \mathcal{O}(10^{-4})$.

The impact of each different Type of 2HDM is very similar in the majority of the parameter space we investigate as the top quark loops provide the dominant contributions and each model has the same set of couplings to the top quark. It is only in extreme $\tan\beta$ regions that we observe some differences in the $\chi^2$ from each model, where we see that Types I and Y give the same results, as do Types II and X, owing to these models having the same lepton sectors, indicating that the lepton contributions become significant only when the couplings become extreme; see Table~\ref{tab:xis}. In all scenarios and models, increasing $\cos(\beta-\alpha)$ does increase the sensitivity, but this effect is relatively minor and insufficient to give any meaningful additional exclusion regions.

This lack of sensitivity to 2HDM states in parameter regions not already excluded by other sectors means that the MUonE experiment is unlikely to be able to provide a distinct signal of BSM physics in the Higgs sector, which is of benefit to the aim of the experiment of measuring the leading-order hadronic contribution to the anomalous magnetic moment of the muon. The measurement will be free from potential contamination in the case that a 2HDM of any Type is realised in a fashion consistent with the theoretical and collider bounds.

\subsection{Electroweak Phase Transition}
Previous works \cite{Atkinson:2021eox,Atkinson:2022pcn} assessed the implications of the global fits on Type I and II 2HDMs for a strong first-order electroweak phase transition (SFOEWPT).
The significance of the different Yukawa couplings of different Types on the EWPT in 2HDMs is minimal, so the main difference in the results from the four flavour-conserving 2HDMs comes from the parameter constraints of the fits. 
For a EWPT to be of strong first-order, it is favoured that there is a sizeable difference between $m_{H^+}$ and $m_{H^0}$; it is sufficient for $m_{A^0}\sim m_{H^+}$. 
From the theoretical constraints found in \cite{Atkinson:2021eox}, one can see that this then inhibits larger Higgs masses from producing the SFOEWPT since the larger the mass scale, the more degenerate the new masses become. 
Favouring masses $\lesssim 1\,$TeV in order to generate a sufficient mass separation, we find that, similarly to Type I, $m_{H^+}$ still has enough freedom in Type X in order to generate this SFOEWPT within $1\sigma$ of our best fit.

However in Type Y, $m_{H^+}$ is most constrained out of all Types and has a similar lower bound to that of Type II.
Furthermore, within $2\sigma$ $\tan\beta$ is constrained to be much larger than in other Types; typically a SFOEWPT will require a $\tan\beta\sim{\cal O}(10)$.
The constraints on both $\tan\beta$ and $m_{H^+}$ together prohibit a SFOEWPT from being realised within $2\sigma$ of our global fits for Type Y. 
Beyond $2\sigma$, the constraints on $\tan\beta$ significantly relax, however the constraints on $m_{H^+}$ still remain stronger than those of the other Types.
While these mass constraints reduce the allowed range of masses to test for a SFOEWPT, they still are sufficiently low enough to generate a mass splitting which, along with a low value of $\tan\beta$ allowed above $2\sigma$, can generate a SFOEWPT. 
Considering a wide range of parameters at various $\sigma$ from our best fit, we conclude that a SFOEWPT generated from the Type Y 2HDM would currently lie at or above $3\sigma$ from our best fit.

\section{Conclusions}
\label{sec:conc}
The search for new physics beyond the Standard Model is a high priority of the particle phenomenology programme across many different experimental settings. The sizeable tension of the anomalous magnetic moment of the muon could be a concrete pointer towards a more comprehensive theory of particle physics. 
The occurrence of this tension specifically highlights the lepton sector as a BSM source, which, when understood in the context of two Higgs doublet extensions motivates a combined analysis of collider and flavour physics data for 2HDM scenarios X and Y. 
To our knowledge this has been presented here comprehensively for the first time. 
We find that each of these analyses exclude large amounts of the parameter space of the models, with the combination of the two improving upon this further as the flavour and collider constraints exhibit a good degree of complementarity. 

Within the 2HDM-I and 2HDM-X we can accommodate the measurements 
of the muon anomalous magnetic moment by BNL/Fermilab using as SM prediction the White Paper result~\cite{Aoyama:2020ynm}.
The hadronic vacuum polarisation contribution is known to be a relevant source of uncertainty for the BSM interpretation of the muon anomalous magnetic moment measurement.
Recent lattice results~\cite{Borsanyi:2020mff,Ce:2022kxy,Alexandrou:2022amy}
 show tension with the data-driven methods employed to formulate the SM 
 expectation consensus~\cite{Aoyama:2020ynm}. Whilst future updates can be expected here, this discrepancy can be addressed experimentally at 
 MUonE, which provides a precise data-driven extraction of the vacuum 
 polarisation. Given the high sensitivity of MUonE to this naively 
 subdominant contribution, it is important to also understand the impact 
 of lepton-philic new physics for this experiment and the resulting impact on the interpretation of MUonE results in terms of the hadronic vacuum polarisation contribution. This becomes particularly relevant for the 2HDM-X and Y scenarios as they can address the anomaly in some parameter regions that are consistent with experimental 
 findings~\cite{Athron:2021iuf}. In this work, we provide the first comprehensive analysis of electroweak radiative corrections for these 
 scenarios at MUonE in conjunction with constraints from flavour collider observables (also including the partonic vacuum polarisation contribution consistently at leading order). We find that these electroweak corrections do not play a constraining role for the parameter region favoured by flavour and collider data. Consequently, MUonE data can be fully interpreted as a measurement of the hadronic vacuum polarisation contribution in these scenarios without the need to correct for BSM effects.

In passing, 
we found that all four 2HDM Types can reproduce large shifts 
in $m_W$, as indicated by the recent CDF measurement and we identify the
possibility of having a SFOEWPT  within $1\sigma$ of our best fit point
 in 2HDM-X and at or above $3\sigma$ of our best fit point in 2HDM-Y.

\medskip\noindent{\bf{Acknowledgments}} --- 
We are grateful to Christine T.H. Davies, Rusa Mandal, Panagiotis Stylianou and Thomas Teubner for helpful discussions. We thank Stephan Hessenberger for making available the 2-loop $m_W$ computation included in Sec.~\ref{sec:2HDM_mW}.
O.A. is funded by a STFC studentship under grant ST/V506692/1. The work of M.B. is supported by Deutsche Forschungsgemeinschaft (DFG, German Research Foundation) through TRR 257 ``Particle Physics Phenomenology after the Higgs Discovery''. C.E. is supported by the STFC under grant ST/T000945/1, the Leverhulme Trust under grant RPG-2021-031, and the Institute for Particle Physics Phenomenology Associateship Scheme. Parts of the computations carried out for this work made use of the OMNI cluster of the University of Siegen.
\bibliography{references} 
\end{document}